\documentclass[trackchanges,twocolumn]{aastex7}

\newcommand{\fink}{{\sc Fink}}

\begin{document}

\title{Strategy for identifying Vera C. Rubin Observatory kilonova candidates for targeted gravitational-wave searches}

\author[orcid=0000-0002-6100-537X]{Simon Stevenson}
\altaffiliation{These authors contributed equally}
\affiliation{Centre for Astrophysics and Supercomputing, Swinburne University of Technology, Hawthorn, VIC 3122, Australia}
\affiliation{OzGrav, ARC Centre for Excellence of Gravitational Wave Discovery, Hawthorn, VIC 3122, Australia}

\email[show]{spstevenson@swin.edu.au}  

\author[orcid=0000-0001-8211-8608]{Anais M\"{o}ller}
\altaffiliation{These authors contributed equally}
\affiliation{Centre for Astrophysics and Supercomputing, Swinburne University of Technology, Hawthorn, VIC 3122, Australia}
\affiliation{OzGrav, ARC Centre for Excellence of Gravitational Wave Discovery, Hawthorn, VIC 3122, Australia}
\email[show]{amoller@swin.edu.au} 

\author[orcid=0000-0002-1357-4164]{Jade Powell}
\affiliation{Centre for Astrophysics and Supercomputing, Swinburne University of Technology, Hawthorn, VIC 3122, Australia}
\affiliation{OzGrav, ARC Centre for Excellence of Gravitational Wave Discovery, Hawthorn, VIC 3122, Australia}
\email{jpowell@swin.edu.au}  

\begin{abstract}
Since the observation of the binary neutron star merger GW170817 and the associated kilonova AT2017gfo, the next joint gravitational-wave/optical kilonova has been highly anticipated.
Overlapping observations between the Vera C. Rubin Observatory and the international gravitational-wave detector network are expected soon.
Wide-field survey facilities, such as Rubin, can serve dual roles in gravitational-wave astronomy: conducting dedicated optical counterpart searches following gravitational-wave triggers and, through surveys such as the Legacy Survey of Space and Time (LSST), providing opportunities for fortuitous kilonova discoveries during routine operations.
We use simulations to develop a strategy for identifying kilonova candidates observed by Rubin and processed by the \fink\ broker. 
These candidates can be used as astrophysical triggers for a targeted gravitational-wave search.
We simulate kilonovae light-curves for the first year of Rubin with the latest observing strategy for the Wide-Fast-Deep and the Deep Drilling Fields. 
Assuming a kilonova rate of 250\,Gpc$^{-3}$\,yr$^{-1}$, we find that Rubin should observe $\sim 40$ kilonovae per year within the gravitational-wave detector horizon ($\sim 500$\,Mpc). However, only $\sim 4$ will be received by the brokers with a signal-to-noise ratio larger than 5.
Most of these will be faint, and detected 1--2 days following the neutron star merger. 
Photometric and spectroscopic follow-up will be limited to large telescopes.
Using archival data from the Zwicky Transient Facility (ZTF) and our proposed selection criteria, we estimate a minimum contamination of at least 30 events per month from other transients and variables, even under our strictest selection criteria.
A deep gravitational-wave search targeting Rubin kilonova candidates may lead to the next multi-messenger discovery.
\end{abstract}



\section{Introduction} 

The detection of gravitational waves (GWs) from the binary neutron star merger GW170817 \citep[][]{Abbott:2017PRLGW170817} was a landmark discovery in astronomy. 
As well as being detected in gravitational waves, the event was detected across the electromagnetic spectrum, from radio to gamma-rays, with the short gamma-ray burst GRB 170817a detected $\sim 2$\,s following the merger \citep[][]{Abbott:2017ApJGW170817Multimessenger,Abbott:2017ApJGW170817GRB170817A}.
The precise sky localisation allowed for the discovery of an optical counterpart, 
AT2017gfo, associated with the kilonova \citep[][]{Abbott:2017ApJGW170817Multimessenger,Coulter:2017Science,Smartt:2017Nature,Arcavi:2017Nature,Andreoni:2017PASAGW170817Australian,Soares-Santos:2017ApJL,Tanvir:2017ApJL}, allowing for GW170817 to be localised to its host galaxy, NGC 4993, at a distance of only 40\,Mpc \citep[][]{Abbott:2017ApJGW170817Multimessenger}.

A second binary neutron star merger, GW190425 \citep[][]{Abbott:2020ApJGW190425}, was detected during the third observing run \citep[O3;][]{Abbott:2021PhRvXGWTC2,Abbott:2023PhRvXGWTC3} of the International Gravitational Wave Detector Network (IGWN), consisting of Advanced LIGO \citep[][]{Aasi:2015CQGAdvancedLIGO}, Virgo \citep[][]{Acernese:2015CQGAdvancedVirgo} and KAGRA \citep[][]{Akutsu:2019NatAstKAGRA,KAGRA:2021PTEP}. 
However, due to the poor sky localisation and larger distance compared to GW170817, no associated counterparts were found.
The fourth LIGO-Virgo-KAGRA observing run is currently ongoing. Initial projections suggested that up to $10^{+52}_{-10}$ binary neutron stars may be detected (\citealt[][]{Abbott:2020LRRObservingScenarios}, see also \citealt[][]{Petrov:2022ApJ}, \citealp{Andreoni:2022ApJS} and \citealt{Kiendrebeogo:2023ApJ}), but no significant binary neutron star candidates have been detected in low-latency in O4 to date\footnote{see \url{https://gracedb.ligo.org/superevents/public/O4/} for gravitational-wave alerts from O4}. 
\citet{Niu:2025arXiv} present a potential sub-threshold binary neutron star merger in O4 \citep[see also][]{Abac:2025GWTC-4}, though it was not discovered until almost two years after the gravitational-wave data were taken, which makes multi-messenger follow-up difficult. 
Detecting additional binary neutron star mergers with associated kilonovae\footnote{In this paper, we restrict the discussion to binary neutron star mergers as progenitors to kilonovae. In principle, some fraction of neutron star black hole (NSBH) mergers may also produce gamma-ray bursts and kilonovae. However, mass ratios close to unity and high black hole spins are required to tidally disrupt the neutron star. Predictions based on the current constraints predict that only a small fraction of NSBH will tidally disrupt \citep[e.g.,][]{Biscoveanu:2023MNRAS}, and so we choose to ignore this class throughout this paper. Much of the discussion would generalise trivially to also include NSBH binaries. } will allow for more precise constraints on cosmology, including measuring the Hubble constant \citep[][]{Abbott:2017Nature,Abbott:2023ApJ,Abac:2025GWTC4Cosmology}.

Three main kilonova discovery channels are possible: 
(i) Target of Opportunity (ToO) observations following up gravitational-wave events (or gamma-ray bursts) using wide-field telescopes, 
(ii) detection of orphan kilonovae in wide-field optical surveys such as Rubin and the subsequent search for low-signal-to-noise GW detections and 
(iii) detection of orphan kilonovae with no associated gravitational wave transient.
The former two connect GW detectors with multi-wavelength ones.

Many strategies have been proposed for ToO observations to follow up binary neutron star merger candidates based on low-latency alerts from gravitational-wave detectors \citep[e.g.,][]{Antier:2020MNRAS,Andreoni:2022ApJS,Ahumada:2024PASP}. 
A common issue for follow up to gravitational-wave alerts is the large sky localisations of typical gravitational-wave events \citep[][]{Abbott:2020LRRObservingScenarios,Petrov:2022ApJ,Kiendrebeogo:2023ApJ} which are difficult to cover with small field of view instruments, along with the large distances (compared to GW170817), making the prospects of detecting faint kilonovae even more challenging (though see \citealp{Rastinejad:2022Nature} and \citealp{Levan:2024Nature} for kilonova candidates associated with gamma-ray bursts within 350\,Mpc). Alternate methods, such as an optical search for orphan kilonovae, have also been explored \citep[e.g.,][]{Doctor:2017,Scolnic:2018,Setzer:2019MNRAS,Andreoni:2019PASP,Andreoni:2020,Aivazyan:2022,Melo:2024,vanBemmel:2025MNRASKNTRAP,Fulton:2025MNRAS}.

The Vera C. Rubin Observatory will soon enter full science operations with the Legacy Survey of Space and Time \citep[LSST;][]{Ivezic:2019ApJ}. 
Rubin is scheduled to begin science operations in October, 2025, coinciding with the final few months of the fourth observing run of LIGO-Virgo-KAGRA \citep[O4;][]{Abbott:2020LRRObservingScenarios}.
Rubin will image the Southern sky with a large field of view (almost 10 deg$^2$) and is capable of imaging the sky to a limiting magnitude of $r < 24.5$ with a $30$\,s exposure. 
As part of LSST, it will survey the visible Southern sky every few days, revolutionising transient astronomy. 
However, this survey strategy is not ideal for rapidly evolving transients such as kilonovae \citep[cf.][]{Andreoni:2022ApJSRubinCadence,Andrade:2025PASP}, which have a peak brightness of $M_{AB} = -15$ magnitude and decline by up to a magnitude per night in $r$ band (and even faster in the bluer bands, assuming similar intrinsic behaviour to AT2017gfo), leading to only a few detections before the kilonova fades, making classification challenging. 
In addition, kilonovae are rare, and Rubin will produce millions of detections per night, consisting of supernovae, luminous red novae \citep[][]{Howitt:2020MNRAS}, flare stars, variable sources, asteroids and kilonovae. 
This means there will be significant contamination for kilonovae searches \citep[e.g.,][]{Cowperthwaite:2018ApJ,Fulton:2025MNRAS,Barna:2025arXiv}.
Due to the large volume of data being produced by Rubin, bespoke brokers such as \fink\ \citep[][]{Moller:2021MNRASFINK} have been developed, providing a way to process the data and classify transients.

Gravitational waves from nearby binary neutron star mergers (within the current binary neutron star range; see Section~\ref{subsec:detection_rate_GW})  will be detected by low-latency automated search pipelines \citep[][]{Chaudhary:2024PNAS}, as was the case for GW170817 \citep[][]{Abbott:2017PRLGW170817}. 
More distant binary neutron star mergers, with lower signal-to-noise ratios, may not be detected in online searches, but produce triggers in more sensitive offline searches \citep[e.g.,][]{Magee:2019ApJL,Abbott:2024PRDGWTC2.1}.

In this paper, we develop a strategy for identifying kilonova candidates observed by Rubin and processed by \fink.
These candidates can be used as astrophysical triggers to perform a targeted gravitational-wave search in LIGO/Virgo/KAGRA data. 
This work extends initial estimates done with early Rubin LSST observing strategies \citep{Setzer:2019MNRAS} and includes difference image processing effects. 
By utilising the known location and approximately known time of a kilonova candidate, a  targeted search for gravitational waves can be more sensitive than a standard all-sky search \citep[cf.][]{Kochanek:1993ApJL,Kelley:2013PRD}, boosting the detection prospects for a joint gravitational-wave kilonova candidate by 40--100\% \citep[][]{Kelley:2013PRD,Williamson:2014PRDPyGRB}. 
This opens up the possibility of detecting sub-threshold binary neutron star mergers that may otherwise be missed in all-sky searches.
In order to conduct such a search, a list of kilonova candidates needs to be compiled. We explore strategies for selecting kilonova candidates, estimate the number and properties of kilonovae detected by \fink/Rubin and estimate the level of contamination expected in such a sample. 

This paper is organised as follows: 
In Section~\ref{sec:gw_searches_for_BNS} we review existing search strategies for GWs from binary neutron star mergers. 
In Section~\ref{sec:rubinandkne} we introduce the Vera C. Rubin Observatory and the Legacy Survey of Space and Time (LSST). 
We discuss our simulations of kilonova lightcurves in Section~\ref{subsec:knesimulations}.
In Section~\ref{sec:identifying_kilonovae_FINK} we introduce \fink\ and perform simulations of kilonova light-curves with the latest Rubin observation strategy to estimate the population of kilonovae that will be detected, and highlight the significant contamination expected from other astrophysical transients and variables.
We also devise a selection method for potential orphan kilonovae.
In Section~\ref{sec:targeted_subthreshold_GW_search} we discuss our findings in the context of a proposed search for gravitational waves from kilnovoae detected by Rubin.  
Finally we conclude in Section~\ref{sec:conclusions}.

\section{Searches for gravitational waves from binary neutron star mergers}
\label{sec:gw_searches_for_BNS}

The LIGO-Virgo-KAGRA collaborations perform searches for gravitational waves from compact binary mergers in both a low-latency mode (with a goal of identifying a gravitational-wave transient as early as possible following merger, or even potentially prior to the merger; \citealt[][]{Magee:2021ApJL}) and a higher latency offline mode, which is typically more sensitive \citep[e.g.,][and references therein]{Abac:GWTC4Methods}.
These searches do not require any external trigger, and search all of the sky, all of the time when data from gravitational-wave detectors is available. 

Targeted searches are also performed for a broad range of external astrophysical triggers from other messengers that may be associated with the emission of gravitational waves, such as gamma-ray bursts \citep[][]{Abbott:2021ApJLVKGRBO3a,Abbott:2022ApJLVKGRBO3b} and fast radio bursts \citep[][]{Abbott:2023ApJLVKFRBO3a}.
These external messengers typically identify the time and location of a potential source of gravitational waves that can be used to restrict the parameter space that is searched for gravitational-wave signals. 
The reduced parameter space allows for a more sensitive search \citep[][]{Kelley:2013PRD}.

For example, \citet{Williamson:2014PRDPyGRB} find that in a targeted search for gravitational waves coincident with well-localised gamma-ray bursts, the distance to which signals could be detected was increased by a factor of up to 25\%. Since the detection rate scales with the cube of the distance to which signals can be detected, this potentially leads to a doubling of the detection rate.
In this paper, we suggest adopting a similar strategy to target kilonova candidates identified in data from the Rubin Observatory.

\subsection{Binary neutron star merger rate and sensitive volume}
\label{subsec:detection_rate_GW}

The merger rate of binary neutron stars has been empirically measured from a variety of observations including short gamma-ray bursts, Galactic double neutron star binaries observed via radio pulsar timing and gravitational-wave observations of binary neutron star mergers \citep[see][for a review]{MandelBroekgaarden:2022LRRRates}. 
Following O3, the gravitational-wave measurement was $\mathcal{R}_\mathrm{BNS} = 10\text{--}1700$\,Gpc$^-3$\,yr$^{-1}$ \citep[][]{Abbott:2023PRXGWTC3Populations}. 
\citet{Petrov:2022ApJ} use these constraints to predict the number of BNS detections expected during O4 and O5, finding 9--112 and 60--600 respectively.
Recently, analysis of the first part O4 has allowed the binary neutron star merger rate to be constrained to $7.6$--$250$\,Gpc$^{-3}$\,yr$^{-1}$ \citep[][]{Abac:2025GWTC-4Populations}.

Binary neutron stars mergers with a given set of parameters (masses, spins, location etc) produce a strain in a gravitational-wave detector. 
The signal-to-noise ratio ($\rho$) measures the amplitude of the signal strain relative to the noise in the detector. 
We assume that signals are detected if $\rho > \rho_\mathrm{thresh}$, assuming a single detector threshold of $\rho_\mathrm{thresh} = 8$ \citep[][]{Abadie:2010CQGRates,Stevenson:2015ApJ}.
The maximum distance that an optimally oriented and positioned binary neutron star merger is detectable to with $\rho = 8$ is known as the \textit{horizon distance}. 

The binary neutron star inspiral \textit{range} is defined as the angle-averaged distance at which a canonical binary neutron star signal (with two non-spinning neutron stars of mass $1.4$\,M$_\odot$) can be detected. 
During O3, the inspiral range for both LIGO detectors was approximately 100--140\,Mpc \citep[][]{Abbott:2021PhRvXGWTC2,Abbott:2023PhRvXGWTC3}. 
In O4a, the range was increased to around 160\,Mpc \citep[][]{Abac:2025GWTC-4}. 
At the time of writing, during O4c, the range is around 160\,Mpc\footnote{\url{https://gwosc.org/detector_status/O4c/}}.
The range is a factor of approximately 2.26 smaller than the horizon distance \citep[][]{Abadie:2010CQGRates}, giving a horizon distance of $\sim 350$\,Mpc.

\section{Rubin and Kilonovae simulations}
\label{sec:rubinandkne}

\subsection{Rubin LSST}\label{sec:rubin}

The Vera C. Rubin Observatory is located in Cerro Pachón in Chile. Over the next ten years it will perform the Legacy Survey of Space and Time (LSST) which is designed to enable four primary science drivers: a census of the Solar System, a detailed map of the Milky Way, systematic exploration of the time-domain sky, and stringent constraints on dark matter and dark energy \citep{Ivezic:2019ApJ}. Rubin LSST will obtain imaging of the Southern Sky in six band-passes, $ugrizy$, and limiting magnitudes above $24$\,mag with nominal $30$\,s exposures.

The survey is dominated by the Wide-Fast-Deep (WFD) component, which accounts for roughly 90\% of the observing time, with the remainder allocated to a set of mini-surveys targeting particular science cases and a survey of 5 Deep Drilling Fields (DDFs)\footnote{\url{https://survey-strategy.lsst.io/}}. The WFD component scans most of the Southern Sky with at least two nights in a single week and two visits in different filters separated by 30 minutes in a given night. The DDF component targets roughly a couple of times per week the same five selected fields: ELAIS-S1, the XMM Large Scale Structure field (XMM-LSS), Extended Chandra Deep Field South (ECDFS), COSMOS and Euclid Deep Field South (Euclid). In this work we focus on both the WFD and DDF components for kilonova searches.

LSST will revolutionize time-domain astronomy by uncovering an unprecedented number of astrophysical transients, ushering in a new era of optical big data. Variable sources will be detected through a difference-imaging pipeline, with transient candidates publicly reported within 60 seconds via a real-time alert stream. This system is projected to generate on the order of $10$ million alerts per night \citep{ldm612}.

\subsection{Kilonova light-curve simulations}
\label{subsec:knesimulations}

In order to estimate the population of kilonovae that will be observed by Rubin/LSST, we perform a series of simulations of kilonovae lightcurves.
Simulations were generated using SNANA \citep{Kessler:2009} within the PIPPIN framework \citep{Hinton:2020}. Throughout this work, we use SNANA calibrated flux (FLUXCAL) defined from the magnitude with a fixed zeropoint given by: $mag = 27.5 - 2.5 \log_{10}$(FLUXCAL).

For our simulations we use two types of kilonova light-curve models: the Kasen model \citep{Kasen:2017} used also in the {\sc PLAsTiCC} classification challenge \citep{Kessler:2019}, and the Bulla model \citep{Bulla:2019} which was used in the {\sc ELAsTiCC} classification challenge. Kasen models are time-series spectral energy distribution models for binary neutron star mergers, parametrized by ejecta mass, ejecta velocity, and lanthanide fraction. 
Bulla models are based on radiation-transport Monte Carlo code models, and the emission includes two components of the  kilonova ejecta: a high-opacity, lanthanide-rich component and a low-opacity, lanthanide-free component. These models have been compared to the observed emission of AT 2017gfo. Other extensions of this project, outside the scope of this work, would be to incorporate additional models such as the ones in \cite{Sarin:2024}.

\begin{figure*}
\plotone{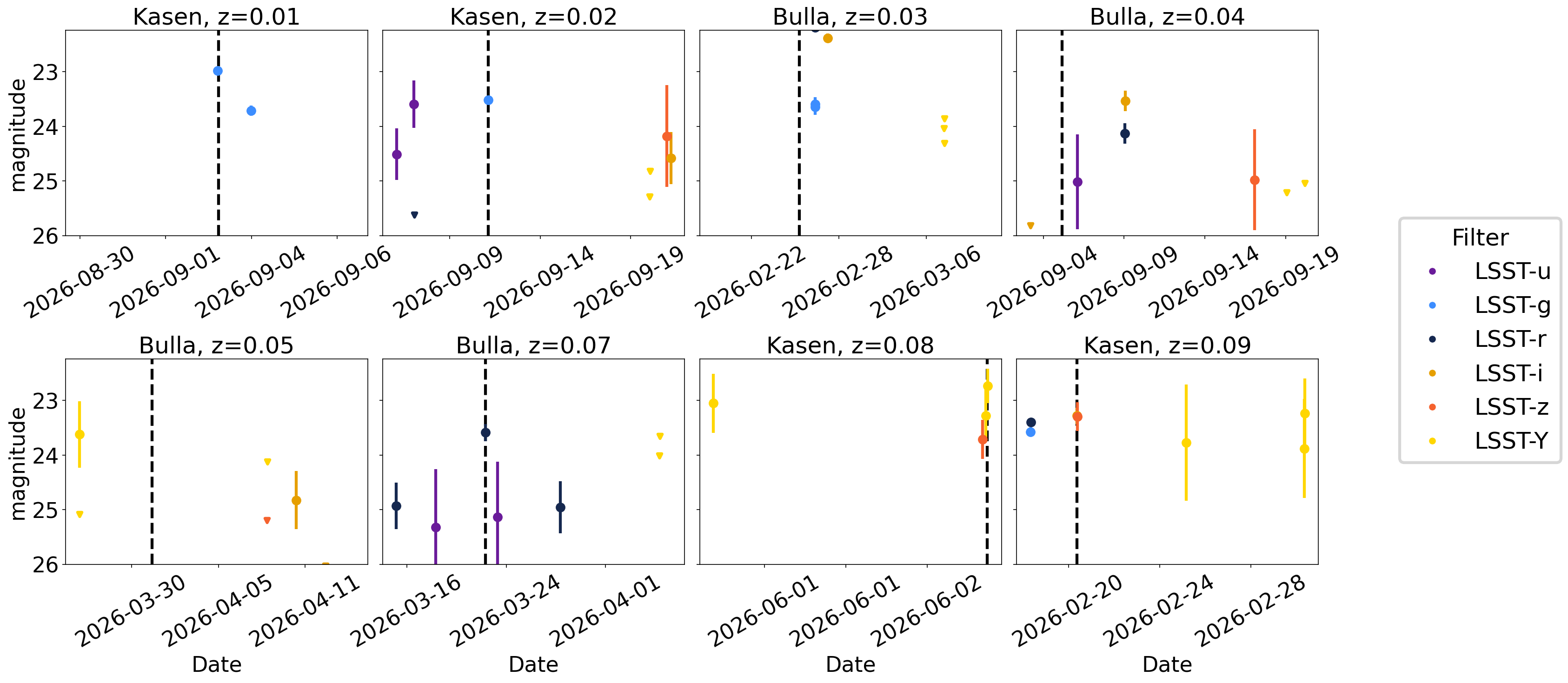}
\caption{Examples of simulated kilonova light-curves using Rubin observing strategy v5.0. We show examples both for Kasen and Bulla models at different redshifts (z). Circles indicate SNR$>3$ and lower triangles SNR$<1$ measurements for difference image photometry. A black vertical line indicates the simulated kilonova peak.}
\label{fig:lcs_sims}
\end{figure*}

We generate one year of LSST light-curves in $ugrizy$ for each kilonova model. We repeat the simulation 100 times, equivalent to producing 4,261 light-curves per model.
We use the observation strategy baseline v5 Ocean\footnote{\url{https://community.lsst.org/t/announcing-new-ocean-ddf-survey-strategy-simulation/10162}} which contains both WFD and DDF. The difference between version 4.3 and Ocean is only in the DDFs which results in minimal changes in our results. We co-add exposures within $30s$ in the simulation to mimic the alert stream of Rubin.
We generate kilonova light-curves in a redshift range of $0.015 < z < 0.11$. Lightcurves are simulated from 30 days before the peak to 60 days after the peak (in the rest frame) to capture any variability in the lightcurve on long timescales.
When simulating kilonova light-curves we assume a rate of $250$\,Gpc$^{-3}$\,yr$^{-1}$ with no redshift dependence \citep[][]{Abac:2025GWTC-4Populations}. In Figure~\ref{fig:lcs_sims} we show examples of generated light-curves for different models at various redshifts. We find that on average Rubin will observe $42\pm4$ kilonovae per year with this strategy. However, as shown in Section ~\ref{subsubsec:alertstream} only a small percentage will have detections with signal-to-noise larger than 3 or 5.

\section{Identification of kilonova candidates in Rubin with Fink}
\label{sec:identifying_kilonovae_FINK}

In this work, we aim to create a search strategy for kilonova candidates directly from Rubin data communicated to brokers. In Section ~\ref{sec:fink}, we introduce \fink\ which is one of the Rubin Community brokers, that aims to enable a broad range of science cases  \citep[][]{Moller:2021MNRASFINK}. In Section~\ref{sec:simalerts}, we translate the simulations introduced in Section~\ref{subsec:knesimulations} into the data that the brokers will receive and study the kilonova candidate properties. We then devise different selection criteria for kilonova candidates in Rubin and test them using archival data from the Zwicky Transient Facility (ZTF) in Section~\ref{sec:Fink_filtering}.


\subsection{Fink}\label{sec:fink}

The \textsc{Fink} community broker is designed to process large-scale time-domain alert streams, filtering, aggregating, enriching, and redistributing data in real time \citep[][]{Moller:2021MNRASFINK}. Built as part of the Rubin Observatory Community Broker ecosystem, \textsc{Fink} was conceived to handle the full alert stream of the Vera C. Rubin Observatory LSST \citep{Ivezic:2019ApJ}. Its modular architecture allows independent science teams to develop filtering and classification routines (“science modules”) for specific astrophysical phenomena. Once integrated, these modules operate within the centralized broker infrastructure, making enriched alert data accessible through its web portal\footnote{\url{https://fink-portal.org/}} and data transfer services\footnote{\url{https://fink-portal.org/download}}.

Since late 2019, \textsc{Fink} has been operating on the ZTF public stream \citep{Bellm_2018}, serving as a precursor to LSST operations. ZTF alerts that pass basic quality criteria are processed through science modules, which add value through catalog cross-matching, statistical analyses, and machine-learning classification routines\footnote{\url{https://fink-broker.readthedocs.io/en/latest/broker/science_modules/}}. The breadth of implemented modules reflects the diversity of the \textsc{Fink} community, spanning supernovae \citep{Leoni:2022, Moller:2020, Moller:2025}, kilonovae \citep{Biswas:2023A&AFinkKilonova}, tidal disruption events \citep{LlamasLanza:2025}, gamma-ray bursts \citep{Masson:2024}, microlensing \citep{Ban:2025}, solar system objects \citep{Lemontagner:2023}, hostless transients \citep{Pessi:2024}, and multi-class classifiers \citep{Fraga:2024}, among others. Each module is developed independently, but their outputs are publicly available, with transparent assumptions to enable informed community use. 

At present, the broker processes $\sim$200,000 ZTF alerts per night in real-time. \fink\ is prepared to scale up to 10 million alerts once LSST starts.

\subsection{Kilonova light-curve simulations in the Rubin alert stream}\label{sec:simalerts}

In this section we translate the kilonova lightcurve simulations introduced in Section~\ref{subsec:knesimulations} into the data that brokers such as \fink\ will receive and study the kilonova candidate properties, including their photometric properties and the typical time delay between neutron star merger and first optical detection.  

\subsubsection{Rubin alert stream}
\label{subsubsec:alertstream}

Rubin LSST will produce an alert stream. The alert stream will contain detections from the Difference Image Analysis pipeline with signal-to-noise (SNR) greater than 5  \citep{ldm612}. For a given object, defined by an event in the same coordinates, if two alerts (SNR$>5$) are emitted in different nights, the latest alert will also contain the history of forced photometry\footnote{\url{https://community.lsst.org/t/alert-information-contained-with-the-two-or-three-visits-per-night/10068/} and \url{https://ldm-612.lsst.io/}}. We show examples of a kilonova light-curve as seen by the broker, and compare to the full detected Rubin light-curve in Figure~\ref{fig:lightcurves}.

\begin{figure}
    \centering
    \includegraphics[width=\linewidth]{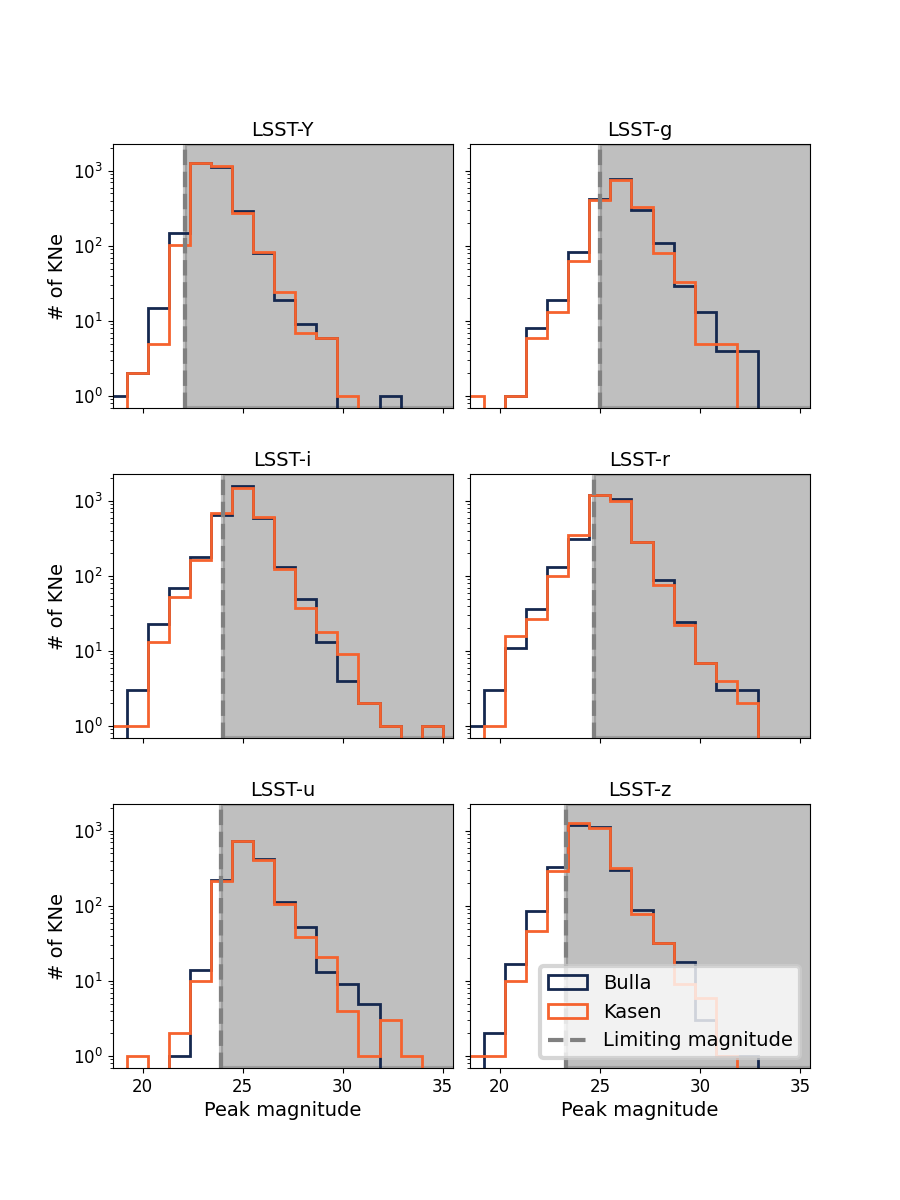}
    \caption{Magnitude histograms for simulated Kilonovae with the Bulla (Blue) and Kasen (Orange) models. We show the nominal 5-sigma source detection limits for each band-pass for Rubin as a vertical line, and shade the region for which photometry is fainter in light grey. 
    }
    \label{fig:magnitude_histograms}
\end{figure}

In Figure~\ref{fig:magnitude_histograms}, we show histograms of the peak magnitude of our simulated kilonova population in each of the LSST filters. Many of the kilonova generated in the volume have peak magnitudes below the 5-sigma point-source depth for the nominal Rubin 30 second exposure.

\begin{figure}
    \centering
    \includegraphics[width=\linewidth]{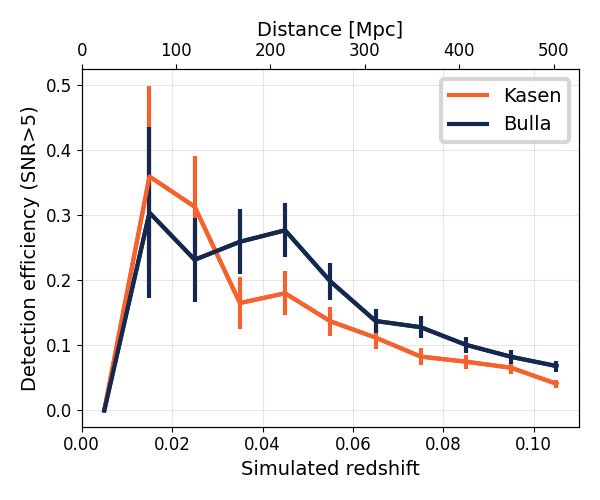}
    \caption{Broker detection efficiency as a function of simulated redshift. We show the ratio of light-curves with an emitted alert ($\mathrm{SNR} > 5$) as a fraction of all simulated light-curves. Results are generated using  two different kilonova lightcurve models, Bulla (blue) and Kasen (orange). The error bars are from propagated Poisson uncertainties.}
    \label{fig:efficiency_alert}
\end{figure}

Using our simulated light-curves introduced in Section~\ref{subsec:knesimulations}, we estimate the fraction of the generated kilonovae light-curves that are communicated to the Rubin brokers (with a minimum of at least one $\mathrm{SNR} > 5$ detection). We find that 11\%/7\% (for the Bulla and Kasen models respectively) of simulated lightcurves have at least one high SNR detection and are communicated to the brokers. The differences between the Bulla and Kasen are due to the diversity of parameter explored for each model and the emission mechanisms taken into account.
We compute the efficiency of alerts (alert emitted vs simulated kilonova) as a function of redshift in Figure~\ref{fig:efficiency_alert}. 

We find that only $6\%$ (Kasen) and $4\%$ (Bulla) of generated light-curves have at least 2 detections with SNR higher than 5. These percentages are further reduced to $< 2\%$ if we require at least 2 detections in different nights to obtain forced photometry.

\subsubsection{Rubin kilonova properties}
\label{subsec:properties}

We find that the two most-common band-passes where kilonovae are detected in Rubin are $i$ and $r$ followed, with half of detections by the filter $z$, and then with small percentages in the order $y,g,u$. This is surprising as previous works showed that $g$ and $i$ were the most probable filters \citep{Andreoni:2022ApJS} for kilonova detection. 
We attribute this difference to the use of updated observing strategies for Rubin (version 4.3 and Ocean) in this work. 

The Rubin Science Verification strategy, shaped through a strong community effort\footnote{\url{https://docs.google.com/document/d/1sssH6buVEfXj7lEOnckuXC4sg-JxNlSlgcKnRffb0fk/edit?usp=sharing}}, has been recently refined\footnote{\url{https://sitcomtn-005.lsst.io}} to prioritize template acquisition in the $i$ band, particularly in regions of the sky where the LIGO-Virgo-KAGRA network has maximum antenna sensitivity \citep[cf.][]{Chen:2017ApJ}. Consequently, the $i$ band is expected to play a central role in the identification of kilonova candidates. 

Given the importance of $i,r$ and $z$ band-passes for the kilonova detection in Rubin, we estimate potential colors which would characterize candidates. We estimate the color of the light-curve around the peak if two band-passes have alerts received by the broker. The dispersion of colors is large with maximum values around $2 ~ mags$. Focusing in $i-r$ we find a smaller dispersion with mean $-0.34 \pm 0.27$ and $-0.22 \pm 0.46$ for Bulla and Kasen models respectively.

We also inspect the rate of decay as a potential indicator of a kilonova light-curve. We highlight that these rates would be useful for a very small percentage of kilonovae given the lack of high SNR detections shown in the last Section. For $i,r,z$ band-passes we find medians of $0.3 \pm 0.4$, $0.4 \pm 0.3$, $0.2 \pm 0.4$ $mag/day$ respectively. When splitting by model for $i,r$, we find the Kasen models have a rate closer to $0.3$ and Bulla models $0.6$ $mag/day$. However, all these rate measurements have very high dispersions.

\begin{figure*}
\gridline{
  \fig{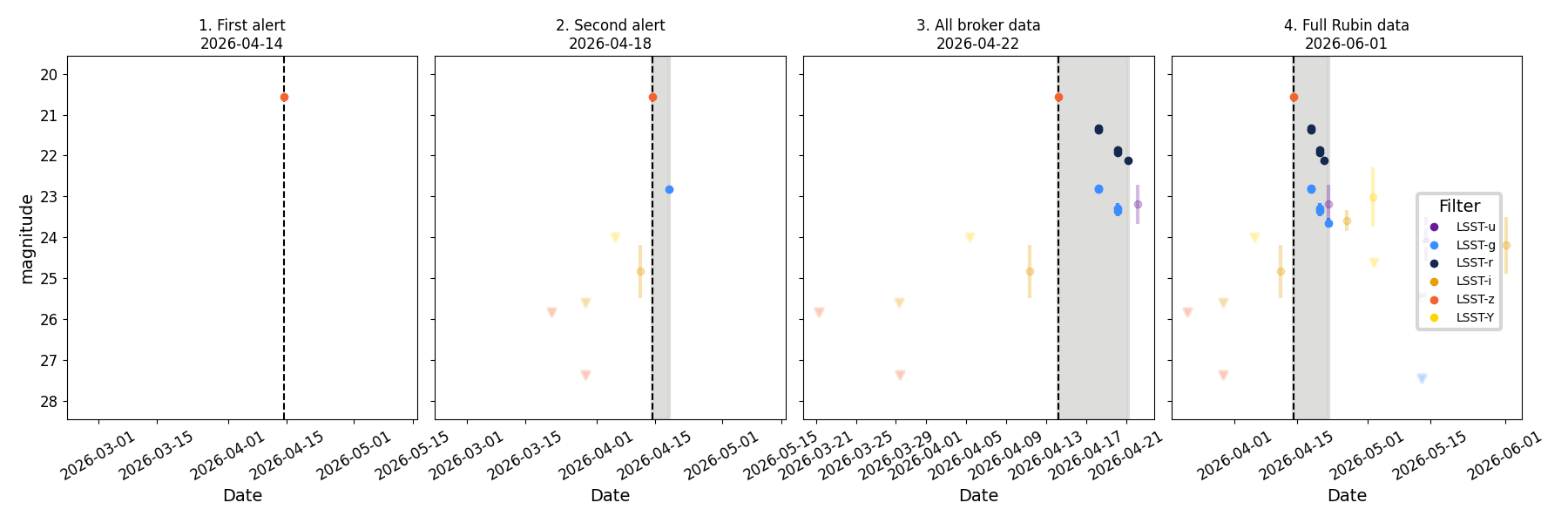}{0.99\textwidth}{(a)}
}
\gridline{
  \fig{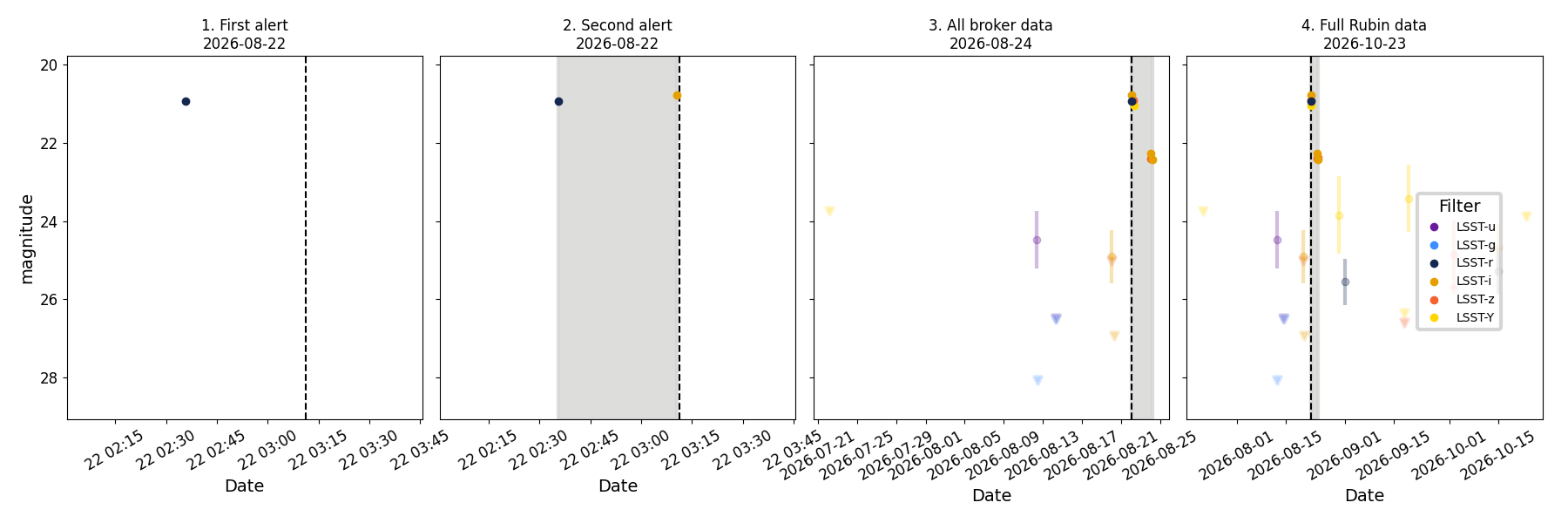}{0.99\textwidth}{(b)}
}
\gridline{
  \fig{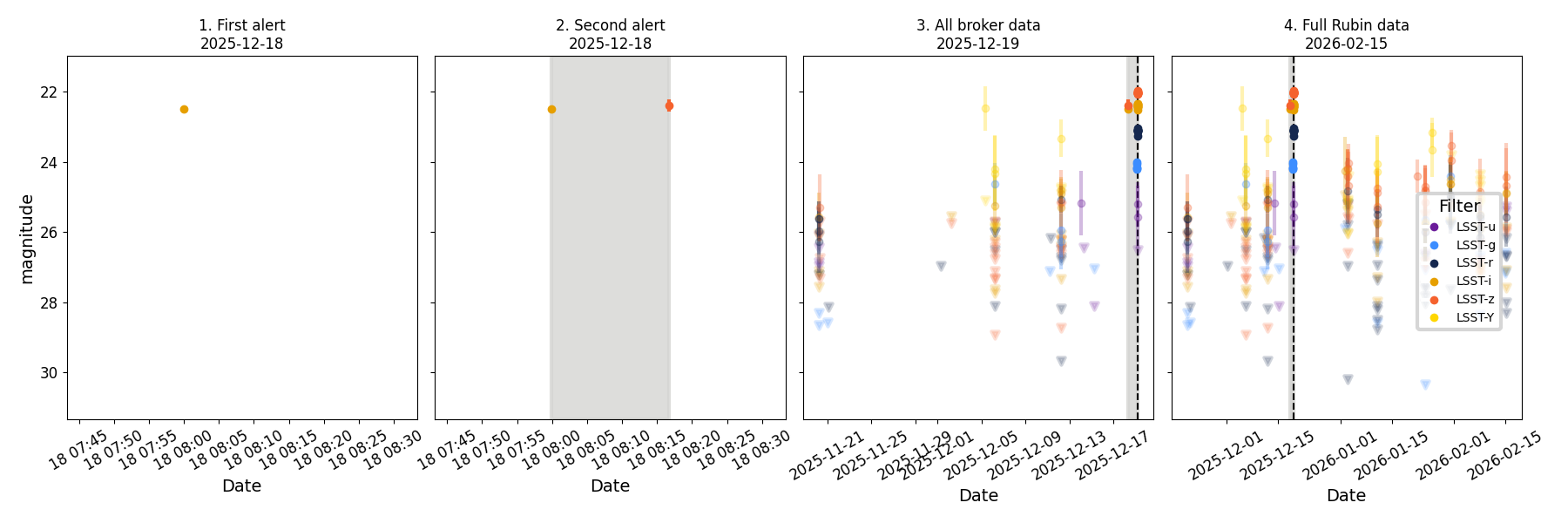}{0.99\textwidth}{(c)}
}
\caption{Well sampled simulated kilonova light-curves for the Kasen model (a) and Bulla model (b) in the Wide Fast Deep fields and (c) the Bulla model in the Deep Drilling Field. 
The first three columns show the data that the \fink\ broker will receive. The different columns show: 1. the first alert received by the broker, 2. the second alert plus forced photometry if the new alert has been received more than one day after the first alert, 3. all data received by the broker, 4. the full light-curve detected by Rubin (including proprietary data). 
Detections with SNR higher than 5 are shown in opaque circles while SNR between 1 and 3 are semi-transparent and are equivalent to the forced photometry received by the broker. We indicate the time of maximum brightness of the kilonova with a black grey line and the range of data received by the broker in grey.}
\label{fig:lightcurves}
\end{figure*}

For the first year of Rubin LSST, and thus for searches of low SNR gravitational-wave detections during O4, the observing strategy of the DDF will be very distinct from that of future years. 
The most updated observing strategy shows a mix of short sequences (2 filters, 2 visits each) at nearly daily cadence, in addition to long sequences (hundreds of visits spread over $griz$) during deep observing seasons. We show a DDF light-curve in panel (c) of Figure~\ref{fig:lightcurves}. This sampling will allow much better constraints on the kilonova modelling using the light-curve. However, we highlight that from our simulations only 5\% (with the observing strategy 4.3, 4\% when using Ocean) of the broker received light-curves are in Rubin DDF.

\begin{figure*}
\gridline{\fig{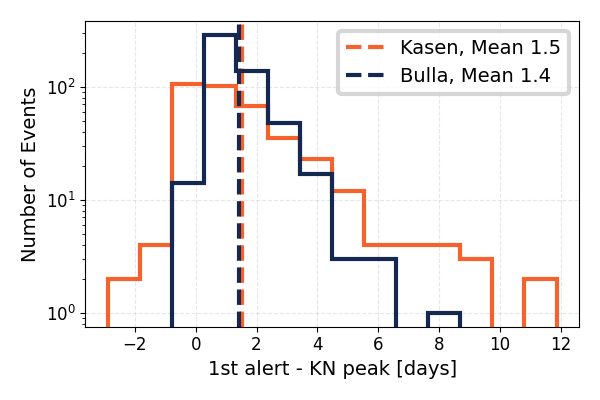}{0.5\textwidth}{(a)}
\fig{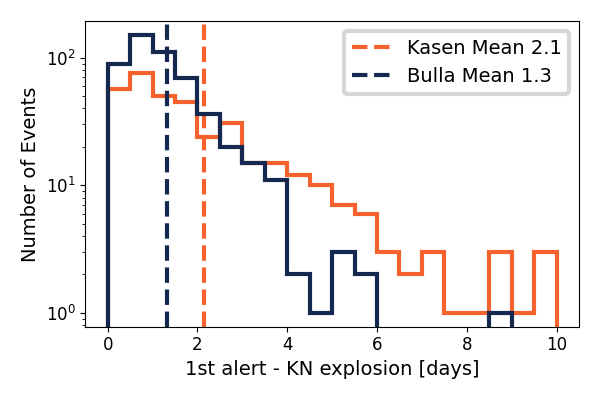}{0.5\textwidth}{(b)}}
\caption{Distribution of duration between the first detection by the broker and: (a) simulated kilonova peak or (b) simulated explosion time. Most kilonovae are detected $\approx 1.5$ days after maximum light and between one and two days after explosion.}
\label{fig:time_between_firstalert_and_peak}
\end{figure*}

\subsubsection{Time between explosion and first detection}
\label{subsec:deltaTime}

Figure~\ref{fig:time_between_firstalert_and_peak} shows the distribution of times between when a kilonova candidate is first detected by Fink/Rubin and a) the time of maximum light and b) the time when the neutron star merger occurred (`explosion time').
We find that most kilonovae are detected around 1.5 days after maximum light and between 1-2 days after explosion.
This figure provides us with an estimate of the typical length of data that would have to be searched in order to be confident that the searched data includes the time of the gravitational-wave emission. 
This is significantly longer than the on-source windows of a few seconds that are used in targeted gravitational-wave searches for gamma-ray bursts \citep[][]{Abbott:2021ApJLVKGRBO3a,Abbott:2022ApJLVKGRBO3b}, and is more similar to targeted searches for core-collapse supernovae \citep[][]{Abbott:2020PRDSNO1O2,Marek:2024PRDSNO3,Abac:2025ApJSN2023ixf}. We discuss implications for targeted gravitational-wave searches further in Section~\ref{sec:targeted_subthreshold_GW_search}.

\subsection{Fink filter for Rubin kilonova candidates}
\label{sec:Fink_filtering}

In the following, we propose different \fink\ broker filters to select kilonova candidates. As a guideline, we take the filters created by the \fink\ and GRANDMA Collaborations \citep{Antier:2020MNRAS}. Since 2021, they have developed 3 different kilonova candidate filters based on: (i) Machine Learning algorithm scores \citep{Biswas:2023A&AFinkKilonova}, (ii) Potential host-galaxy cross-match using Mangrove catalogue \citep{Ducoin:2020Mangrove} and (iii) decline rate of the light-curve \citep{Andreoni:2021}. There are also additional criteria to reject asteroids, known variables and transients.
These filters have been deployed in \fink\ and have identified potential candidates  \citep{Aivazyan:2022,Melo:2024}. 

We also take into account the criteria presented in \citet{Scolnic:2018} and \citet{Setzer:2019MNRAS}. The criteria are (1) have two alerts separated by at least 30 minutes to reject asteroids, (2) be observed in at least two filters with ${\rm SNR} \geq 5$ to characterise them, (3) include ${\rm SNR} \geq 5$ observations separated by no more than 25 days to reject long-lived events, (4) have at least one observation of the same location within 20 days prior to the first ${\rm SNR} \geq 5$ detection, and (5) have at least one observation of the same location within 20 days after the last ${\rm SNR} \geq 5$ detection.

In Section~\ref{sec:simalerts}, we have found that the alert brokers do not receive more than a handful of detections for kilonova candidates. Thus, here we propose different filters with sequential cuts that do not make use of machine learning classification scores nor require at least 4 detections in different filters. We expand also the criteria to reject asteroids and other know transients and variables.

Basic cuts:
\begin{enumerate}
    \item Remove all alerts which are potential subtraction artefacts.
    \item Remove all alerts cross-matched within 2'' with known variable stars and transients using \fink\ and the CDS xmatch service. 
   \item Remove all alerts cross-matched with a Gaia source that has a parallax measurement with $\mathrm{SNR}>5$.
   \item Remove all sources cross-matched within 2'' to Variable star catalogues such as the General Catalog of Variable Stars \citep[GCVS;][]{Samus:2017} and the Variable Star Index \citep[VSX;][]{Kholopov:1992}.
    \item Remove all alerts close by to asteroids catalogued in the Minor Planet Catalogue (MPC).
    \item Select alerts outside the Galactic plane by requiring that the Galactic latitude, $b$, is greater than $20$\,degrees ($|b| > 20^{\circ}$).
    \item Select alerts that have detections for less than 20 days.
\end{enumerate}

These are broad cuts that will select new detected objects, as well as avoid catalogued asteroids, variables and transients (including supernovae). However, a large fraction of the remaining alerts may be uncatalogued cataclysmic variables (CVs), supernovae, novae, Fast Blue Optical Transients (FBOTs) and gamma-ray burst afterglows \citep{Barna:2025arXiv}.

We refine these cuts into two main selection families: all sky searches and galaxy-targeted searches as in \citet{Andreoni:2020}.

For the all sky searches we define the following quality samples:
\begin{itemize}
    \item All sky silver: Less than 3 days of detections and discarding any potential satellite or space debris using a tracklet \citep{Karpov:2023}.
    \item All sky gold: same as silver with the addition of discarding high-probability supernovae (high machine learning supernova score if available).
\end{itemize}

For the galaxy targeted searches we define the following quality samples:
\begin{itemize}
    \item Galaxy silver: use the basic cuts and discard potential satellite tracks and add a condition that the event should be within 2'' of a galaxy with known distance measurement.
    \item Galaxy gold: use the same criteria as all sky gold but add a condition that the event should be within 2'' of a galaxy with known distance measurement.
\end{itemize}
In this work we use as a reference the Mangrove galaxy catalogue. However this could be extended to other catalogues, in particular those with photometric redshifts. 

\subsubsection{Test with ZTF alert stream}

To validate our candidate selection methodology, we applied our filtering criteria to alerts processed by the \fink\ broker \citep{Moller:2021MNRASFINK} from the Zwicky Transient Facility \citep[ZTF;][]{Bellm_2018}. ZTF surveys the northern hemisphere with a limiting magnitude of $\approx 21$ and generates alerts for detections $g,r$ band-passes, which are subsequently processed by \fink.

\fink\ currently employs several kilonova identification filters based on machine learning algorithms, light curve decay rates, and catalog cross-matching. Our analysis of existing Fink kilonova filters over a one-year period yielded an average of 30 candidates per month.

We then applied the filtering pipeline described in Section~\ref{sec:Fink_filtering} to one year of ZTF alert data. This analysis provides both a lower bound for expected Rubin filter performance (given ZTF's lower alert rate) and insights into the contamination sources we anticipate for Rubin candidates.

Table~\ref{tab:filtering_stats} summarizes the filtering statistics for each selection criterion. The basic cuts eliminate approximately 98\% of detected alerts, primarily by matching them to known variable stars, transients, and asteroids. However, the resulting monthly candidate rate remains too high for efficient gravitational-wave follow-up processing. Notably, our analysis of previous years of ZTF data with \fink\ revealed that approximately 50\% of now-cataloged variable stars were initially uncataloged, suggesting that at the start of the survey, Rubin may experience similar or higher stellar contamination rates. Incorporating alert thumbnail analysis could provide additional discrimination against such contaminants.

Among alerts passing the basic cuts, cross-referencing with Fink community filters revealed that while most remain unclassified, approximately 5\% are flagged as supernova candidates and 0.1\% as active galactic nucleus (AGN) candidates.

The All-Sky selection criteria slightly reduce the monthly candidate rate, though supernova and AGN contamination persists. Visual inspection or automated thumbnail-based filtering could help identify and reject centrally located galactic transients.

The Galaxy-targeted criteria substantially reduce candidate numbers by restricting selection to transients in proximity to local volume galaxies. However, incomplete galaxy catalogs may introduce selection biases. Our current analysis utilizes the Mangrove catalog, which is limited to 200 Mpc. The ongoing 4MOST 4-metre Multi-Object Spectroscopic Telescope Hemisphere Survey \citep{Taylor:2023} will provide more complete galaxy samples with spectroscopic redshifts for $z < 0.1$.

We highlight that, given theoretical predictions of longer-duration kilonova signals, we impose no temporal restrictions beyond the 20-day threshold for Galaxy Silver candidates. While Galaxy Silver reduces the overall alert rate, it enables detection of kilonovae with signals extending beyond three days post-merger—a capability that may prove crucial for certain kilonova models.

\begin{figure*}
    \centering
    \includegraphics[width=\linewidth]{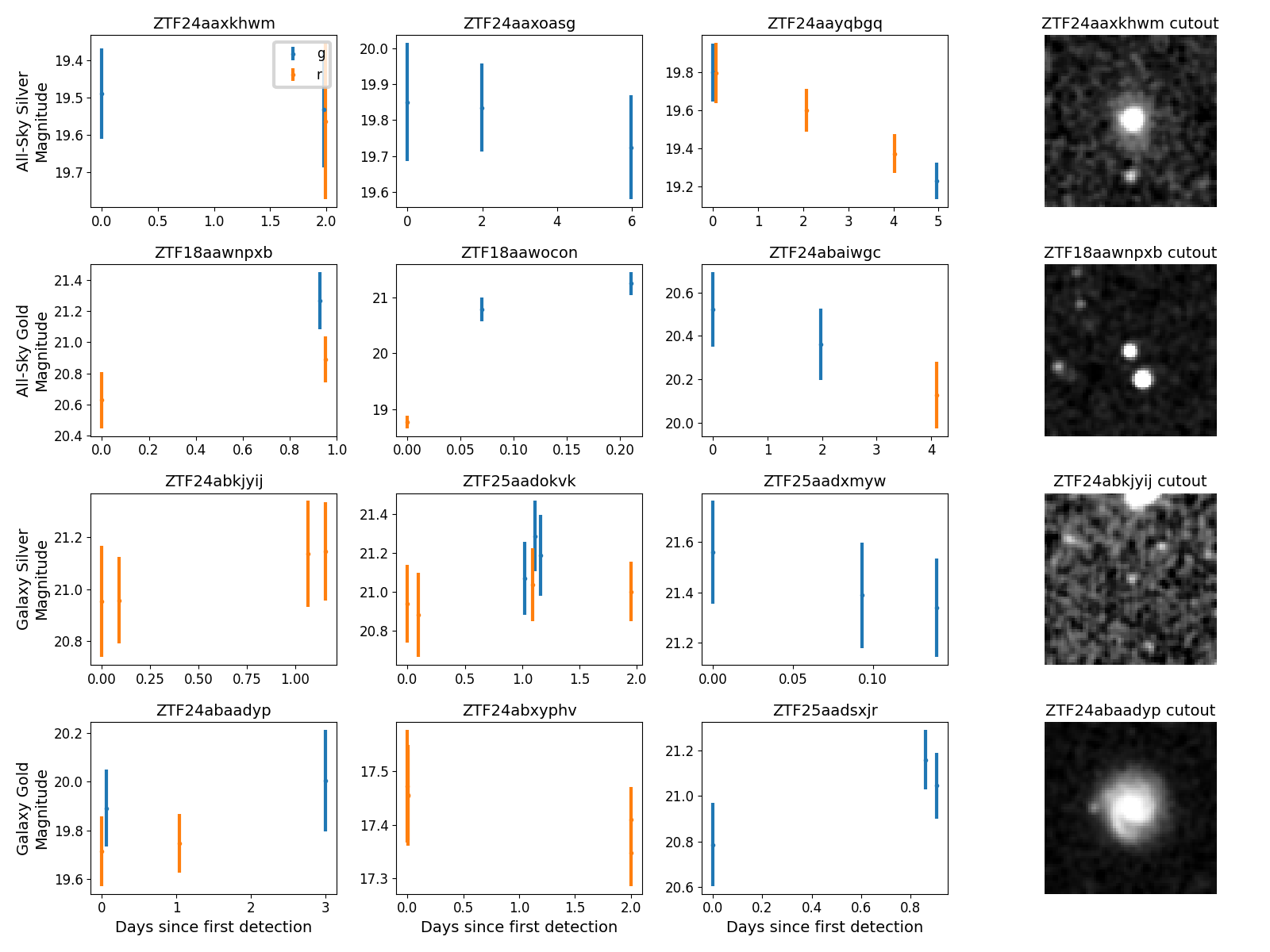}
    \caption{ZTF candidates that pass the different cut stages in the \fink\ broker. We show light-curves for these candidates in the first three columns and science cutouts (no subtraction) in the fourth column. Each row represents one of the sample cuts proposed in Section~\ref{sec:Fink_filtering}.}
    \label{fig:fink_ztf_lcs}
\end{figure*}

Both All-Sky Gold/Silver and Galaxy Gold criteria restrict candidates to fewer than 35 detections per month for ZTF alert volumes and depth. For Rubin LSST this may represent a lower threshold. Together with the filtering, in particular at the beginning of the survey, it will be crucial to perform visual inspection of the candidates, in particular the thumbnails which can show previous sources pointing towards variable stars, events located in the centre of galaxies pointing towards AGNs, as well as other artefacts from image subtraction. We show some examples of candidates that pass our various sample cuts in Figure~\ref{fig:fink_ztf_lcs}.

\begin{table}[h!]
\centering
\caption{Filtering Statistics for 1-year of ZTF data and cuts proposed in Section~\ref{sec:Fink_filtering}. Columns indicate: sample selection, number of events in that sample, efficiency indicated as the percentage of alerts from the 1-year ZTF data and rate per month.}
\label{tab:filtering_stats}
\begin{tabular}{l|r|r|r}
\toprule
Sample & \# events & Efficiency (\%) & Per Month \\
\hline
1-year ZTF & 43,221,791 & 100\% & 3,601,816 \\
Basic Cuts & 53,022 & 0.123\% & 4,418 \\
All-Sky Silver & 51,237 & 0.119\% & 4,270 \\
All-Sky Gold & 49,948 & 0.116\% & 4,162 \\
Galaxy Silver & 379 & 0.001\% & 32 \\
Galaxy Gold & 346 & 0.001\% & 29 \\
\end{tabular}
\end{table}

\subsection{Extending the kilonova search with Rubin proprietary data}\label{sec:Rubinproprietary}

The alert stream is public and contains all events with at least one detection above SNR 5. In this stream, we expect approximately 10 million detections per night of variable and transient sources. In this Section, we explore the use of proprietary data from Rubin LSST which will include lower SNR detections as well as forced-photometry\footnote{\url{https://lse-163.lsst.io}}.

Here we estimate the potential completeness of a kilonova sample when varying the SNR thresholds for a detection in Rubin independent of the contamination by other sources. We vary the SNR threshold needed for a kilonova detection in Table~\ref{tab:lc_detections} and report the fraction of kilonovae with at least one detection above each threshold. Clearly, lowering the threshold increases the number of recovered kilonovae. We note that even when lowering the threshold, the percentage of kilonovae with two detections is small and it is further reduced when requiring more than one day of detections.

\begin{figure}
\gridline{\fig{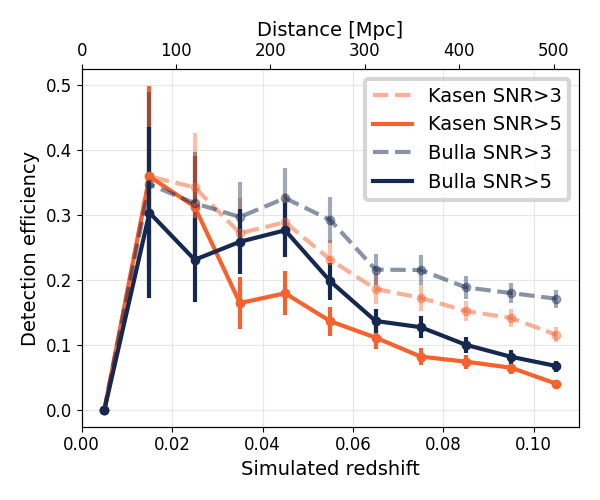}{\columnwidth}{(a)}}
\gridline{
\fig{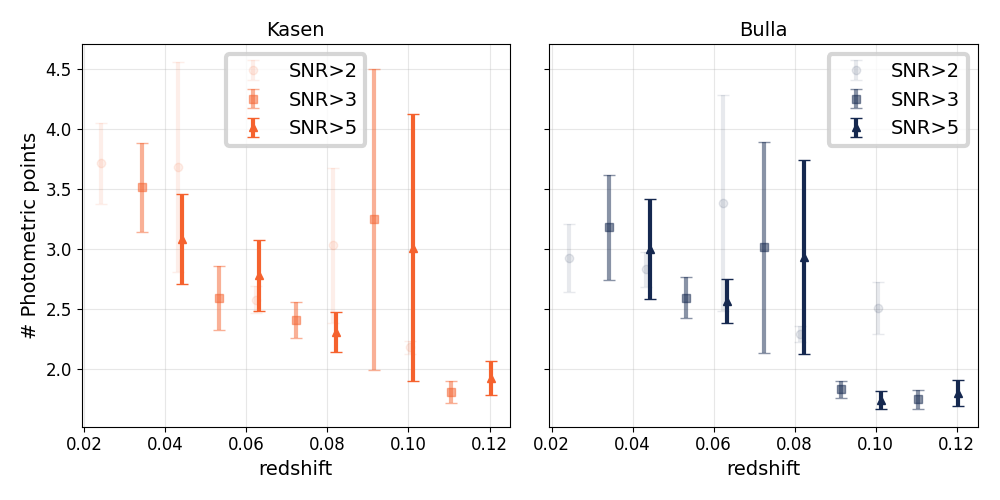}{\columnwidth}{(b)}}
\caption{Kilonova photometry statistics as a function of SNR threshold used for detection.
(a) Detection efficiency of simulated kilonovae for two SNR thresholds as a function of distance (both redshift and Mpc). (b) Average number of photometric points per kilonova in each redshift bin for different SNR thresholds.}
\label{fig:effsnr}
\end{figure}

We further assess the recovery efficiency for a single detection at a given SNR as a function of redshift in Figure~\ref{fig:effsnr} (a), finding improvements of up to 10\% in specific distance bins. In (b) we show that the number of average photometric points per kilonova as a function of redshift does not change substantially as we change the SNR threshold. Forced photometry at the locations of KN candidates may yield additional photometric measurements, particularly in epochs where detections fall below the SNR threshold.

We also examine the number of detections in different band-passes. As with the alert stream, we find that the most common detection filter (whether SNR 5 or 3) is $i$ shortly followed by $r$. This is found for both Kasen and Bulla models.

These findings supports the adoption of a lower SNR threshold for kilonova searches in the Rubin ToO program \citep{Andreoni:2024ToO}. However, for blind searches a compromise between a higher threshold to eliminate spurious detections and the number of detections per light-curve should be considered. 

\begin{table}[h!]
\centering
\caption{Light curve detection statistics for different SNR thresholds. Results are averaged over the two kilonova models, Bulla and Kasen. Percentages are expressed as a fraction of kilonovae with Rubin photometry.}
\begin{tabular}{ccccc}
\toprule
SNR & 1 detection & 2 detections & 2 detections \\ 
& & & different days \\
\hline
1 & 91.8 \% & 80.5 \% & 75.7\%  \\
2 &  57.8 \% & 33.3 \% & 30.0\%  \\
3 &  17.9 \% & 8.9 \% & 4.0 \% \\
5 &  9.4 \% & 5.4 \% & 1.7 \% \\
\hline
\end{tabular}
\label{tab:lc_detections}
\end{table}

\subsection{Optical follow-up scenarios}
\label{subsec:optical_followup}

In this Section, we briefly discuss optical follow-up scenarios for Rubin kilonovae, whether photometric or spectroscopic.

Photometric follow-up to enhance sampling of kilonova light curve evolution represents an optimal strategy. However, the peak magnitudes of the kilonovae in our volume, shown in Figure~\ref{fig:magnitude_histograms}, reveal that many simulated events are quite faint. Current wide-field surveys such as ZTF and LS4 \citep[][]{Bellm_2018,Miller:2025} achieve single-exposure limiting magnitudes of approximately 21.5 mag. Our simulations indicate that only 1–2\% of kilonovae would reach peak brightnesses within these detection limits. Furthermore, given the limited sky overlap between ZTF (in the Northern hemisphere) and the Rubin Observatory (in the Southern hemisphere), this already small fraction would be substantially reduced in practice potentially by a factor larger than 3. Therefore, the single-exposure detection thresholds of the current ZTF alert stream limit its ability to provide substantial additional photometric coverage for the faint kilonova population accessible to Rubin.

Spectroscopic confirmation of kilonova candidates will prove essential for distinguishing between kilonovae and other variables and transients. 
However, given the expected peak magnitudes of kilonova candidates in Rubin, there are substantial observational challenges.
We find that 2m telescopes reaching $19$--$19.5~mag$ \citep[such as the ANU 2.3m telescope equipped with WiFeS;][]{Dopita:2007} would be able to follow less than $0.05\%$ of the candidates. A more pragmatic approach involves targeting the host galaxy of each candidate to determine its redshift, thereby enabling estimation of the absolute magnitude and intrinsic luminosity of the kilonova. This strategy provides crucial context for assessing the physical properties of potential kilonova events.

For real-time kilonova spectroscopy, 8-meter class telescopes will be necessary given the faint nature of these transients. Our analysis demonstrates that approximately 96\% of Rubin-detected kilonovae at redshifts $z < 0.1$ exhibit peak magnitudes fainter than 22 mag, with $\approx 60\%$ between 22 and 24 magnitude. This highlights the substantial observational resources required for comprehensive follow-up programs.

\section{Targeted search for gravitational waves associated with FINK/Rubin kilonova candidates}
\label{sec:targeted_subthreshold_GW_search}

The kilonova candidates identified in Section~\ref{sec:identifying_kilonovae_FINK} provide astrophysical triggers for a targeted search for associated gravitational waves. 
These triggers provide an approximate time of the event, along with a sky location and (if using events with a known host galaxy) distance. 
In a targeted search for associated gravitational waves, this reduces the number of parameters that need to be searched over by up to four, allowing for a more sensitive search to be conducted \citep[e.g.,][]{Kelley:2013PRD,Williamson:2014PRDPyGRB}. This can ultimately lead to additional multi-messenger detections. 

There are several differences between a search for gravitational waves targeting kilonova candidates and one targeting gamma-ray bursts. 
Gamma-ray bursts associated with binary neutron star mergers are expected to occur at almost the exact time of the merger \citep[cf.][]{Abbott:2017ApJGW170817GRB170817A,Abbott:2017ApJGW170817Multimessenger}, providing tight constraints on the time period of data to be searched (of order seconds) for gravitational waves \citep[e.g.,][]{Abbott:2021ApJLVKGRBO3a,Abbott:2022ApJLVKGRBO3b}. 
However, for a kilonova candidate detected by Rubin/\fink, the first detection will likely not correspond to the time of the associated neutron star merger. 
In the worst case scenario, a detection after peak brightness when the transient has already begun to fade, we know the peak must have been some time between the last non-detection and the first detection. 
This provides a window of order the cadence of the survey, which may be up to 5 days.
In Section~\ref{subsec:deltaTime} we showed that most kilonova candidates identified by \fink\ will typically be detected one to two days after the neutron star merger. 
Targeted searches for gravitational waves associated with kilonovae will therefore typically need to search days of data, compared to seconds for searches targeting gamma-ray bursts. 
This leads to a situation that is more similar to targeted searches for gravitational waves from optically detected supernovae, which also typically have on-source windows of several days \citep[e.g.,][]{Abbott:2020PRDSNO1O2,Marek:2024PRDSNO3,Abac:2025ApJSN2023ixf}.
This increases the computational cost of a targeted search, and it is likely that for many events detector data may not exist for the duration that needs to be searched due to the typical duty cycle of $\sim70$\% for gravitational-wave detectors \citep[e.g.,][]{Abac:2025GWTC4Intro}.
We consider the case of GW170817/AT2017gfo as the best case scenario, that is, a kilonova detected before peak brightness, with a well sampled lightcurve, allowing for models to be fit to estimate the time of the merger. 
AT2017gfo was first detected in the optical approximately 12 hours after the merger time.
From the optical data on AT2017gfo alone, the merger time can be estimated approximately \citep[e.g.,][]{Villar:2017ApJGW170817CombinedUVOIR}.

The other main difference between a gravitational-wave search targeting kilonova candidates and targeted searches for gravitational waves associated with gamma-ray bursts is that our list of targets will suffer from significant contamination. The majority of kilonova candidates identified by our selection criteria are not actually kilonovae associated with binary neutron star mergers, but rather other variables or transients such as supernovae and asteroids, with similar colors and magnitudes to kilonovae in one or two detections. The situation of a targeted search with significant contamination is similar to the search for gravitational waves associated with fast radio bursts, where some significant fraction (up to 100\%) of fast radio bursts may originate from events that are not binary neutron star mergers \citep[][]{Abbott:2023ApJLVKFRBO3a}.

\section{Conclusions}
\label{sec:conclusions}

Upcoming contemporaneous observations by the Vera C. Rubin Observatory and the LIGO, Virgo and KAGRA gravitational-wave detectors present an opportunity to detect both kilonovae and gravitational waves from a binary neutron star merger.
We consider the case of a search for subthreshold gravitational waves associated with kilonovae observed by Rubin that occur within the sensitive volume of LIGO-Virgo-KAGRA, in analogy with existing searches for gravitational waves associated with gamma-ray bursts and fast radio bursts \citep{Abbott:2021ApJLVKGRBO3a,Abbott:2022ApJLVKGRBO3b,Abbott:2023ApJLVKFRBO3a}. 
This effort is complementary to efforts to follow-up gravitational-wave triggers by the Rubin ToO programme \citep{Margutti:2018arXiv,Andreoni:2022ApJS,Andreoni:2024ToO}.

In order to carry out such a search in practice, a sample of kilonova candidates is required. 
We perform simulations to estimate the population of kilonovae that will be observed by Rubin within 500\,Mpc ($z < 0.11$), finding that Rubin will detect around 40 kilonovae per year, or 3 to 5 per month. 
However, most of these kilonovae will not have any alerts emitted, as only 10\% will have a detection above $\mathrm{SNR}>5$ for one epoch and will be made public by brokers such as \fink.
This number is reduced to 7\% when requiring two alerts.
Thus the number of strong kilonova candidates is more like 3 per year, in agreement with previous estimates \citep[e.g.,][]{Andreoni:2019PASP,Andreoni:2022ApJS,Ragosta:2024ApJ}.

This work has implications for attempts to perform targeted subthreshold searches for gravitational waves associated with kilonovae using the Rubin LSST public alert stream. 
Our simulations show that typical kilonova candidates detected by Rubin will be detected 1--2\,days following the neutron star merger, requiring several days of gravitational-wave data to be searched.
This is significantly longer than the seconds of data that are typically searched for gravitational waves when targeting gamma-ray bursts and fast radio bursts \citep{Abbott:2021ApJLVKGRBO3a,Abbott:2022ApJLVKGRBO3b,Abbott:2023ApJLVKFRBO3a}. 
This will either require the optimisation of existing search codes, the development of new analyses or the use of suboptimal methods in order to make such a search computationally feasible. 

In addition, we also showed that the sample of kilonvoa candidates detected by Rubin/\fink\ is heavily contaminated by other variables and transients, such as supernovae and asteroids. 
We have developed selection criteria designed to suppress this contamination as much as possible. 
We then estimated the number of contaminants that match our selection criteria using archival data from the Zwicky Transient Facility. 
We find that even using our strictest selection criteria, there will be around 30 events per month matching our selection criteria that are not kilonovae. 
Our estimate provides a lower limit for the contamination rate for Rubin/LSST. The true contamination rate for Rubin is currently unknown. Based on the ratio of the number of alerts emitted by ZTF ($\sim 2 \times 10^{5}$ per night) and Rubin ($\sim 1 \times 10^{7}$ per night), it could potentially be up to a factor of 50 worse than this. 
The true contamination rate from Rubin should be well measured by the start of the Fifth LIGO/Virgo/Kagra observing run. 
The severe contamination will also increase the computational cost of a subthreshold gravitational wave search.
For kilonovoa candidates with a known distance (e.g., from a host galaxy with a known redshift), a binary neutron star merger origin may be able to be excluded if the gravitational wave searches can produce an exclusion distance greater than the distance of the event \citep[cf.][]{Abadie:2012ApJ}.

An example of a recent kilonova imposter during O4 is SN2025ulz (identified by ZTF; \citealt{Stein:2025GCN}), which was initially considered a possible counterpart to the gravitational-wave trigger S250818k due to its color and fade rate \citep[e.g.,][]{Hall:2025arXivEarly}. After approximately five days, the transient began to rebrighten, and was subsequently classified as an unrelated type II supernova \citep[][]{Franz:2025arXiv,Gillanders:2025arXiv,Hall:2025arXivDESI,Kasliwal:2025arXiv,OConnor:2025arXiv,Yang:2025arXiv}. This classification was only possible due to substantial follow-up of this transient.

In contrast to previous analyses of kilonovae in Rubin \citep[e.g.,][]{Andreoni:2022ApJS}, we find that the most common band-passes an alert will be emitted in is $i$ followed shortly by $r$. For programmes such as ToO we encourage reducing the SNR threshold drastically. 
Only 21\% of kilonovae are detected with $\mathrm{SNR}>3$ and up to 50\% with $\mathrm{SNR}>2$. Blind searches can also benefit from joining different alert streams such as ZTF, LSST and LS4 in particular if the surveys are observing the same patches of the sky. This would require joint stream filtering and potentially benefit from multi-modal machine learning models to identify the most promising candidates.

Using the known sky location of the candidate and the approximately known merger time, a more sensitive (`targeted') search for gravitational-wave emission associated with a binary neutron star merger can be performed \citep[][]{Kelley:2013PRD}.
Similar previous targeted searches have shown an improvement of up to 25\% in distance over an untargeted, all-sky search, leading to a potential doubling of the detection rate \citep[][]{Williamson:2014PRDPyGRB}. 
Since Rubin only surveys $\sim 18,000$\,deg$^{2}$ in the Southern hemisphere, we expect that the increase that can be expected is smaller than for all-sky gamma-ray burst monitors.
Further studies will be required to assess the feasibility of our proposal, and determine the improvement that can be realised in practice. 
Targeted searches for gravitational waves associated with kilonova candidates from Rubin may provide the next multi-messenger observation.

\begin{acknowledgments}
We thank Emille Ishida, Julien Peloton and Marion Pillas for useful comments on this manuscript, as well as Rick Kessler for help with SNANA. We thank the annonymous referee for comments that improved the manuscript.

\end{acknowledgments}





%
\facilities{LIGO, Virgo, KAGRA, Rubin, ZTF}

\software{This work made use of the following software packages: \texttt{astropy} \citep{astropy:2013, astropy:2018, astropy:2022}, \texttt{Jupyter} \citep{2007CSE.....9c..21P, kluyver2016jupyter}, \texttt{matplotlib} \citep{Hunter:2007}, \texttt{numpy} \citep{numpy}, \texttt{python} \citep{python}, and \texttt{scipy} \citep{2020SciPy-NMeth} and \fink\ \citep[][]{Moller:2021MNRASFINK,Biswas:2023A&AFinkKilonova}.
}

This research has made use of NASA's Astrophysics Data System.

Software citation information aggregated using \texttt{\href{https://www.tomwagg.com/software-citation-station/}{The Software Citation Station}} \citep{software-citation-station-paper, software-citation-station-zenodo}.

The Vera C. Rubin Observatory is funded by the U.S. National Science Foundation and the U.S. Department of Energy's Office of Science.

This material is based upon work supported by NSF's LIGO Laboratory which is a major facility fully funded by the National Science Foundation.

This work was developed within the \fink\ community and made use of the \fink\ 
community broker resources. \fink\ is supported by LSST-France and CNRS/IN2P3. This research made use of the cross-match service provided by CDS, Strasbourg and the SIMBAS database and has made use of the International Variable Star Index (VSX) database, operated at AAVSO, Cambridge, Massachusetts, USA. This research used resources of the National Energy Research Scientific Computing Center (NERSC).

Notebooks producing the plots in this paper can be found at \url{https://github.com/anaismoller/Rubin_KNe_with_Fink_LVK}.

This research was supported by grants from the Australian Research Council (ARC). 
AM is supported by the ARC Discovery Early Career Research Award (DE230100055). 
SS is supported by an ARC Discovery Early Career Research Award (DE220100241).
Parts of this research were conducted by the ARC Centre of Excellence for Gravitational Wave Discovery (OzGrav), through project number CE230100016.



\bibliography{bib}{}
\bibliographystyle{aasjournalv7}



\end{document}